\documentclass[preprint,12pt]{elsarticle}
\usepackage{graphicx}
\usepackage{amssymb}
\usepackage{lineno}
\usepackage{subfig}
\usepackage{amsmath}
\usepackage{siunitx}
\usepackage{colortbl}

\journal{xx}

\definecolor{Gray}{gray}{0.75}
\definecolor{Blue}{rgb}{0.239,0.522,0.776}
\definecolor{Red}{rgb}{0.878,0.4,0.4}
\definecolor{Green}{rgb}{0.576,0.769,0.49}

\begin{document}
\begin{frontmatter}

\title{Raiders of the lost SAR: Radiofrequency cycles of magnetic nanoflowers inside a tumor}

\author[1]{I.J. Bruvera}

\author[1]{D.G. Actis}
\author[2]{P. Soto}
\author[2]{V. Blank}
\author[2]{L. Roguin}
\author[1]{M. B. Fernández van Raap}
\author[1]{P. Mendoza Z\'elis}
\ead{pmendoza@fisica.unlp.edu.ar}
\address[1]{Instituto de Física La Plata (IFLP),UNLP-CONICET. Departamento de Física, Facultad de Ciencias Exactas, UNLP. La Plata, Argentina}
\address[2]{Instituto de Química y Físicoquímica Biológica (IQUIFIB), Departamento de Química Biológica, Facultad de Farmacia y Bíoquímica, UBA. CONICET, Buenos Aires, Argentina}

\begin{abstract}
Radiofrequency magnetic cycles of \textit{ex vivo} melanoma tumor tissue loaded with Fe$_3$O$_4$ nanoflowers (NF) were measured for several field conditions and compared with the cycles of a ferrogel (FG) obtained incorporating identical NF in agarose gel. Results were studied in order to understand a reported specific power dissipation (SAR) reduction of the NF in the actual application medium (tumor) in comparison with a typical characterization model as the FG. The linearity of the response, together with coercive field and SAR values were analyzed.\\ Additionally, a novel method for the determination of the NF mean relaxation time is presented.
Results show a systematic difference in magnetic response between the NF incorporated in the tumor and those in the FG for all field settings (\{98, 170, 260\} kHz and \{17.4, 54.0\} kA/m) with SAR reductions above 50\%.
\end{abstract}


\end{frontmatter}

\section{Background}
\subsection{Power dissipation of magnetic nanoparticles under radiofrequency field}
The power dissipation of magnetic nanoparticles (MNPs) exposed to radiofrequency fields (RF) enables the development of several applications in biomedicine such as controlled drug release \cite{bruvera2015integrated}, frozen tissue rewarming \cite{manuchehrabadi2017improved} and oncologic thermotherapy    \cite{stephen2021recent}. In all these applications, MNPs absorb energy from the field and release it to their surroundings as heat. The factor of merit for this process is named Specific Absorption Rate (SAR) and is expressed as power dissipation per unit mass of magnetic nanoparticles at a given RF field amplitude and frequency. The SAR value of a set of MNP depends not only on the RF parameters but also on the supporting medium and their spatial distribution, which determines the dipolar interactions between particles.

\subsection{SAR determination methods}
The typical determination of the power dissipation of MNPs under RF consists in the measurement of the adiabatic temperature rise in a liquid suspension of particles known as ferrofluid (FF) or in solid suspensions as ferrogels (FG). This method, where a highly localized temperature probe is used, is simple and direct but presents several drawbacks such as sample instability during measure time ($\sim$ 1-10 min), heat losses of the sample holder, thermal stratification in liquid samples and poor spatial representativity of the temperature measurement in non liquid samples\cite{wang2012potential}. An alternative method based on the inductive measurement of the RF magnetic loop of the sample (ESAR for Electromagnetic SAR) has been developed in the last years \cite{gudoshnikov2013ac,garaio2014}. This inductive method not only allows to perform very fast measurements ($\sim$ 2 s) in all kinds of samples but also provides, together with SAR values, the actual magnetization vs. field cycle of the MNPs.

\subsection{Power dissipation mechanism}
The power dissipation of MNPs exposed to RF originates in the magnetization lag with respect to the applied field. This lag is characterized by the relaxation time of the magnetization, which is determined by two processes. The fluctuation of the magnetic moment between  orientations of easy magnetization within the particle is known as Nèel mechanism. On the other hand,  the rotation of the magnetic moment, together with the particle itself when suspended in a finite viscosity medium is known as Brown mechanism. Nèel relaxation time is determined by MNP magnetic volume and its anisotropy constant.  Brown relaxation depends on medium viscosity and MNP hydrodynamic volume. If the two mechanisms are accessible, the effective relaxation time is a parallel sum of both. The time dependence of the magnetization, and hence, the shape of the magnetization cycle, depends on the effective relaxation time and through it, on the interaction between the MNPs and the medium. SAR value, meanwhile, is proportional to the cycle area.\cite{rosensweig2002heating} So, cycles of different shapes can yield the same SAR value.

In oncologic thermotherapy, MNPs are injected into the tumoral tissue where they are subjected to interactions with the cellular structures and the surrounding medium. So the particles end up in a much different environment  than in the FF where they are usually characterized. Several studies report a noticeable change in the MNPs SAR  when changing the dispersion medium, \textit{i.e}. fluid, solid matrix or actual tumoral tissue.\cite{lahiri2017magnetic, rousseau2021influence} In particular Coral \textit{et al} reports a considerably reduction in SAR for MNPs characterized in a FG when measured in tumor tissue.\cite{coral2018nanoclusters} 

\subsection{Determination of the relaxation time from magnetization cycles}
When a sample is exposed to an alternating magnetic field of the form $H(t) = H_0 \cos (\omega t)$, if $H_0$ is small enough, the resultant magnetization takes the form
$M(t)= H_0 (\chi^{'} \cos(\omega t)+\chi^{''} \sin(\omega t))$
where $\chi^{'}$ s the in-phase component, and $\chi^{''}$ is the out-of-phase component of the susceptibility
$\chi$\cite{rosensweig2002heating}. When the field frequency is high enough, magnetization lags behind the field. In this nonequilibrium situation, the magnetization will tend to relax to the equilibrium value corresponding to the instant value of the magnetic ﬁeld. To describe this process, Shliomis\cite{shliomis1974magnetic} has postulated the relaxation equation

\begin{equation}
    \frac{\partial M(t)}{\partial t}=\frac{1}{\tau}(M_{eq}(t)-M(t))
    \label{schi}
\end{equation}

where  $\tau$ is the effective relaxation time, $M_{eq}(t)$ is the equilibrium magnetization and $M(t)$ is the instant magnetization. For magnetic fields of small amplitude, the response is linear \textit{i.e.} $M_{eq}(t)=H_0 \chi_0^{'} \cos(\omega t)$. In this case,  substituting the expressions of $M_{eq}(t)$ and $M(t)$ in equation \ref{schi} gives 

\begin{equation}
\chi=\frac{\chi_0}{1+i\omega t}    
\end{equation}
and
\begin{equation}
   M(t) = H_0\chi_0 \cos (\omega t-\tan^{-1}(\omega \tau)) 
   \label{mag}
\end{equation}

Then,  both $\chi_0$ and $\tau$ can be determined by fitting the measured magnetization with equation \ref{mag}.\\

In this work we present a comparison between RF cycles of MNPs supported in an agarose gel matrix (typical solid model characterization) and MNPs incorporated into \textit{ex vivo} tumor tissue from a murine melanoma model. The batch  of magnetite MNPs was used to fabricate an agarose ferrogel, and a PBS  FF to be injected \textit{in vivo} into a mouse tumor in order to perform hyperthermia treatment as an evaluation for oncological applications as reported in \cite{coral2018nanoclusters}. Afterwards, the tumor was extracted and its RF cycles measured for several field conditions. The results are compared with those obtained with FG samples.
Additionally, the cycles were used to determine and compare the relaxation time of the MNPs for all samples at every field condition. 

\section{Methods}
\subsection{Tumor and ferrogel samples}
\begin{figure}
    \centering
    \includegraphics[width=0.5\linewidth]{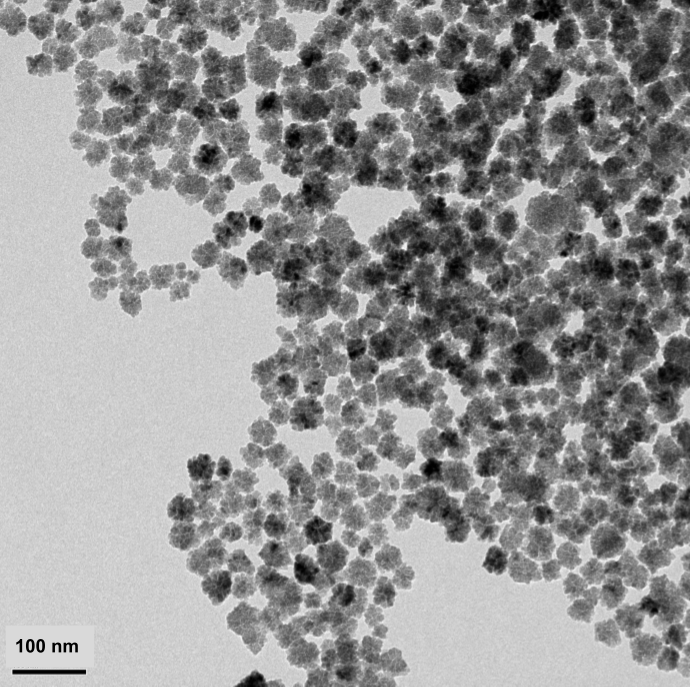}
    \caption{TEM image of magnetic nanoflowers.}
    \label{NF}
\end{figure}

Nanoflower-like magnetic nanoclusters (NF) were used to prepare two composite materials: an agarose gel with homogeneously distributed NF (FG), and several NF loaded tumor samples obtained from mice challenged with B16-F0 melanoma cells.\\    
The nanoflowers (fig. \ref{NF}),  34(4) nm size,  composed of crystallographically aligned magnetite nanoparticles 8(2) nm in diameter, were synthesized by coprecipitation in polyol mixture using a protocol adapted from \cite{hugounenq2012iron} and resuspended in distilled water to a Fe concentration of 3.6 g/L. A hydrodynamic size of 45(14) nm (mean value of number distribution) resulted from dynamic light scattering measurements carried out in diluted aqueous suspension at pH ~7. For VSM measurements, the NFs display superparamagnetic behavior (no coercivity and Langevin type magnetic response), a RT saturation magnetization of  39(1) A/m/kg$_{Fe}$ and a SAR value of 10.3 W/g$_{Fe}$ calorimetrically determined in water suspension at a concentration of 3.6 kg$_{Fe}$/L at RF conditions of 100 kHz and 9.3 kA/m.\\
FG sample was prepared by adding 7 mg of agarose in 700 $\mu$L of the NF aqueous suspension. The temperature of the mixture was increased up to boiling point and was finally cooled down to RT.\\
A melanoma tumor, 320 mm$^3$ volume and 0.410 g mass, resulting from the inoculation of a mouse with B16-F0 cells was injected with 10 mg$_{Fe}$ of NF dispersed in PBS and excised for its study \textit{ex vivo} after 24 h. The tumor was cut into 4 pieces (T1 of 58 mg, T2 of 77.5 mg, T3 of 129.1 mg and T4 of 88.8 mg) after excision. The most external part, named T1, is the closest one to the skin while T4 is the more internal.

\subsection{RF measurements}
The RF field was generated by a power source-resonator set Hüttinger TIG 2,5/300 with a [30; 300] kHz nominal frequency range and a 2.5 kW maximum output reaching a maximum field amplitude of 54 kA/m with a water refrigerated 6 turn coil of 2.5 cm diameter
In order to measure the magnetization of the sample during the application of the RF field, an \textit{ad hoc} device was built as reported in \cite{bruvera2019typical}. In this case,  identical 3D printed transparent PEG capsules were used to hold both the samples and a $Gd_2O_3$ paramagnetic pattern used for calibration.\\  
RF magnetization cycles were obtained for tumor samples T1, T2, T4 (T3 was damaged during measurement), and for FG sample at three frequencies (98(1) kHz, 170(1) kHz and 260(1) kHz) and two field amplitudes: 17.4 kA/m and 54 kA/m (maximum available amplitude). Three measurements were made for each sample at every field condition, each measured value is the  average of 64 cycles. Results are mean $\pm$ SD of three determinations.\\ 
After the measurements, magnetite content of each sample was determined by UV-Vis spectroscopy with a method derived from \cite{adams1995determining}.
\begin{table}
\centering
\begin{tabular}{|c | c|}
\rowcolor{Gray}
\hline
Sample  & [Fe] (g/L) \\
\rowcolor{Blue}
\hline
FG & 4.97(2)\\
\hline
T1 & 1.5(1)\\
\hline
\rowcolor{Red}
T2 & 7.7(2)\\
\hline
\rowcolor{Green}
T4 & 30.9(5)\\
\hline
\end{tabular}
\caption{Fe concentration of every measured sample.}
\end{table}
\begin{figure}
    \centering
    \includegraphics[width=0.8\linewidth]{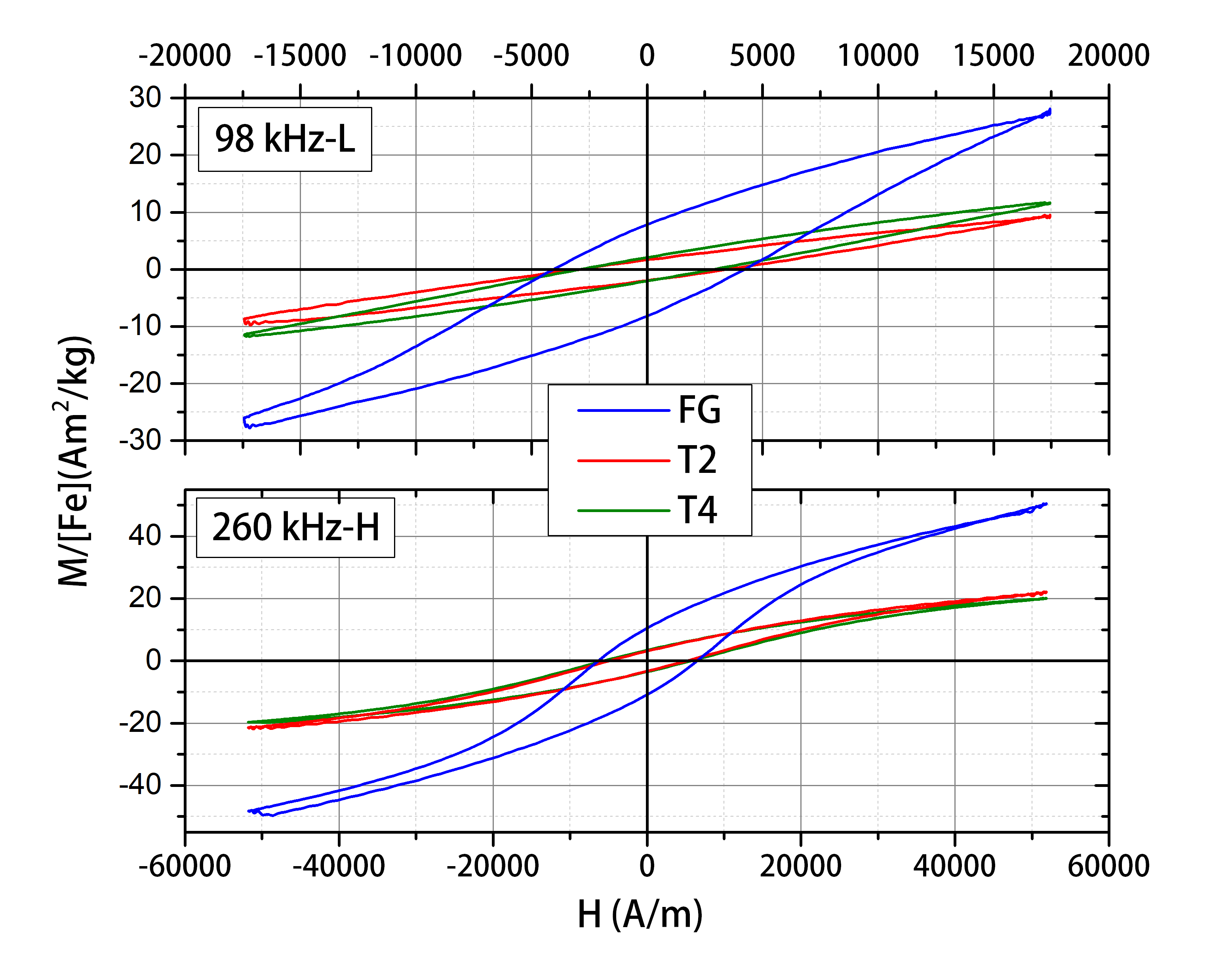}
    \caption{Typical RF cycles for FG (blue), T2 (red) and T4 (green) at low (17.4 kA/m, top panel) and high (54.0 kA/m, bottom panel). Magnetization values are normalized by iron concentration of each sample.}
    \label{ciclos}
\end{figure}
\section{Results}
\subsection{Magnetite concentration}
Table 1 shows the results for iron concentration in FG and tumor   tissue samples. T4 sample has a larger concentration than T2 and T1 indicating a strong inhomogeneity in MNPs distribution in the original tumor. This can be understood as a direct consequence of the local injection of the original ferrofluid . On the other hand, FG sample has a concentration similar to T2.

\subsection{Cycle and SAR comparison}
Figure \ref{ciclos} shows typical results for RF cycles of T2 and T4 tumor portions and FG sample. Magnetization values were normalized by iron content.\\
In general, T2 and T4 samples cycles are compatible with each other and different from FG cycles. Cycles of T1 present a noisy linear magnetic response in correspondence to its low MNP content (not show). Additionally, control tumor samples with no MNP loading were measured showing a weak diamagnetic response typical of hydrated biological tissue (not show). \\ 
\begin{figure}
    \centering
    \includegraphics[width=0.7\linewidth]{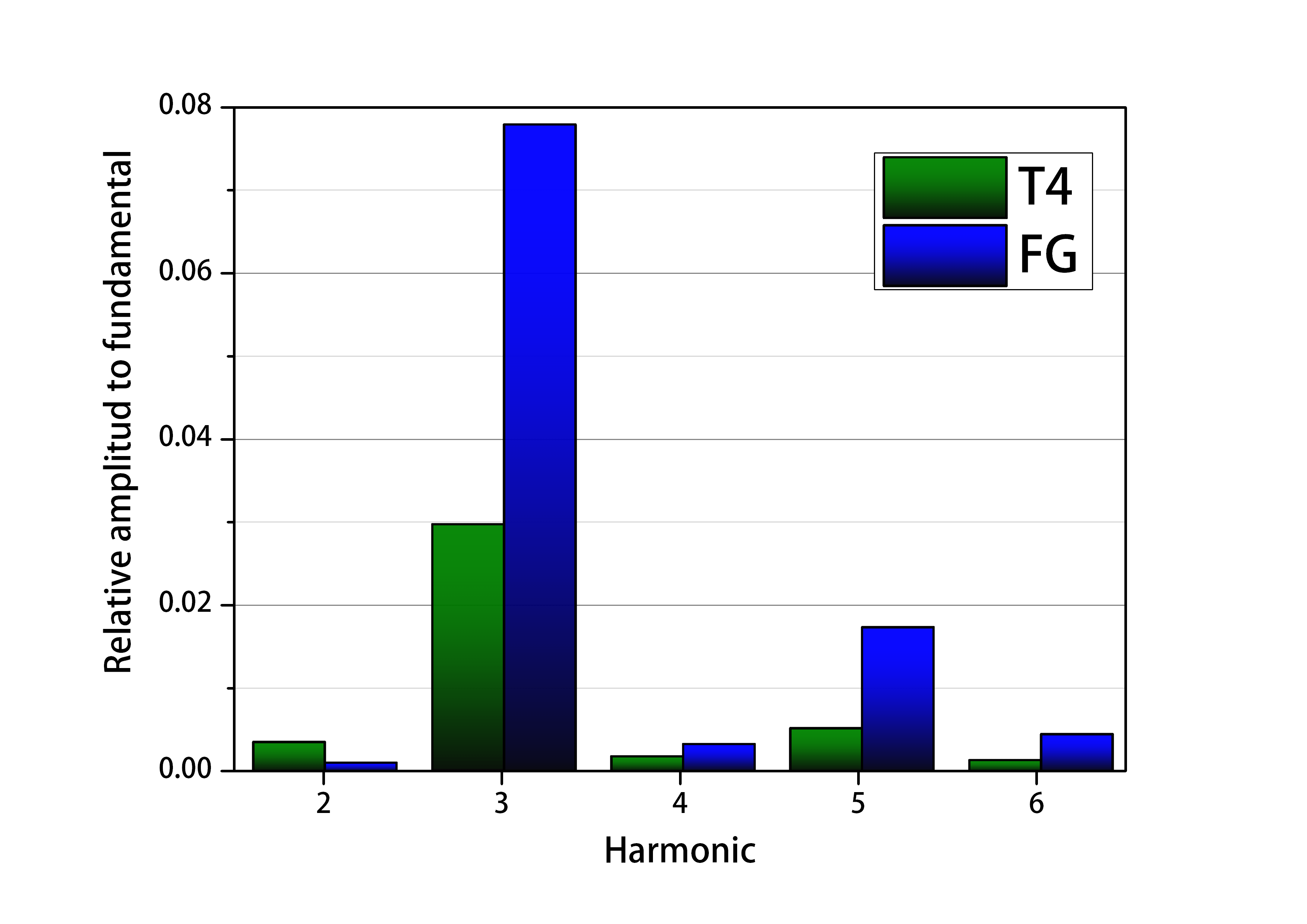}
    \caption{Harmonic components of $M(t)$ normalized to fundamental frequency (1st harmonic) amplitude for FG (blue) and T4 (green) at 17.4 kA/m, 260 kHz RF.}
    \label{FFT}
\end{figure}
For low field amplitude (17.4 kA/m) MNPs in the tumor samples respond linearly as can be noticed from the elliptical shape of the cycles. FG cycles, on the other hand, present a more sigmoidal shape indicating the presence of harmonics of the fundamental field frequency in the magnetization. Figure \ref{FFT} shows a comparison between the harmonic components of $M(t)$ for FG and T4  signals at a low amplitude 260 kHz RF. It can be seen that the first two odd harmonics are present in the FG signal with a bigger amplitude than in the tumor response.\\

\begin{figure}
    \centering
    \includegraphics[width=0.8\linewidth]{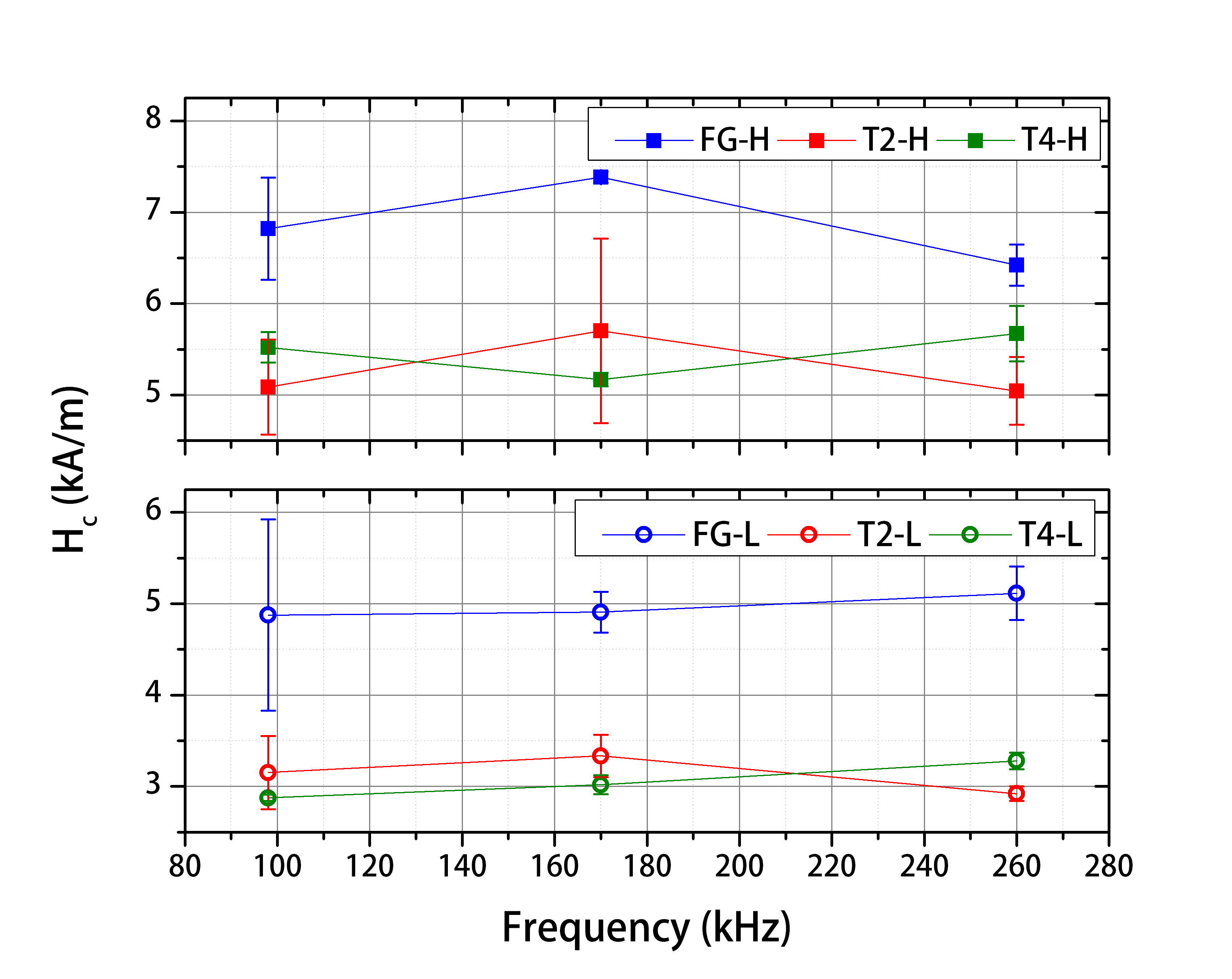}
    \caption{Coercive field $H_c$ vs. field frequency for FG (blue), T2 (red) and T4 (green) for high (H=54 kA/m, top panel) and low (L=17.4 kA/m, bottom panel) field amplitudes.}
    \label{Hc}
\end{figure}

Coercive field $H_c$ is systematically larger for the FG and compatible between tumor samples for all field configurations as shown in figure \ref{Hc}.

\begin{figure}
    \centering
    \includegraphics[width=0.8\linewidth]{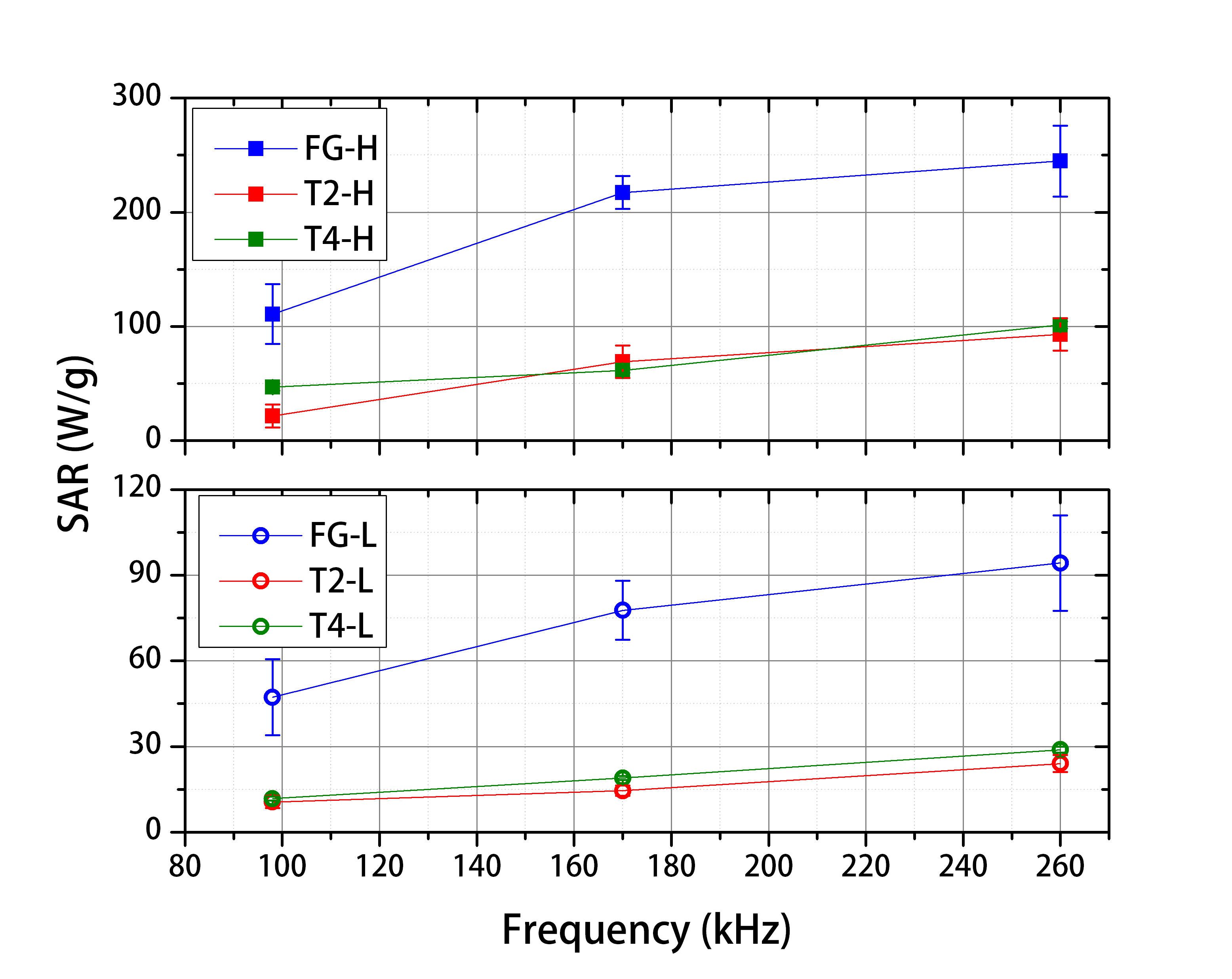}
    \caption{SAR values vs. field frequency for FG (blue), T2 (red) and T4 (green) for high (H=54 kA/m) and low (L=17.4 kA/m)  field amplitudes.}
    \label{SAR}
\end{figure}

SAR values (fig. \ref{SAR}) are systematically larger for FG and almost identical between tumor samples for all field configurations.\\

\subsection{Relaxation time comparison}
\begin{figure}
    \centering
    \includegraphics[width=0.7\linewidth]{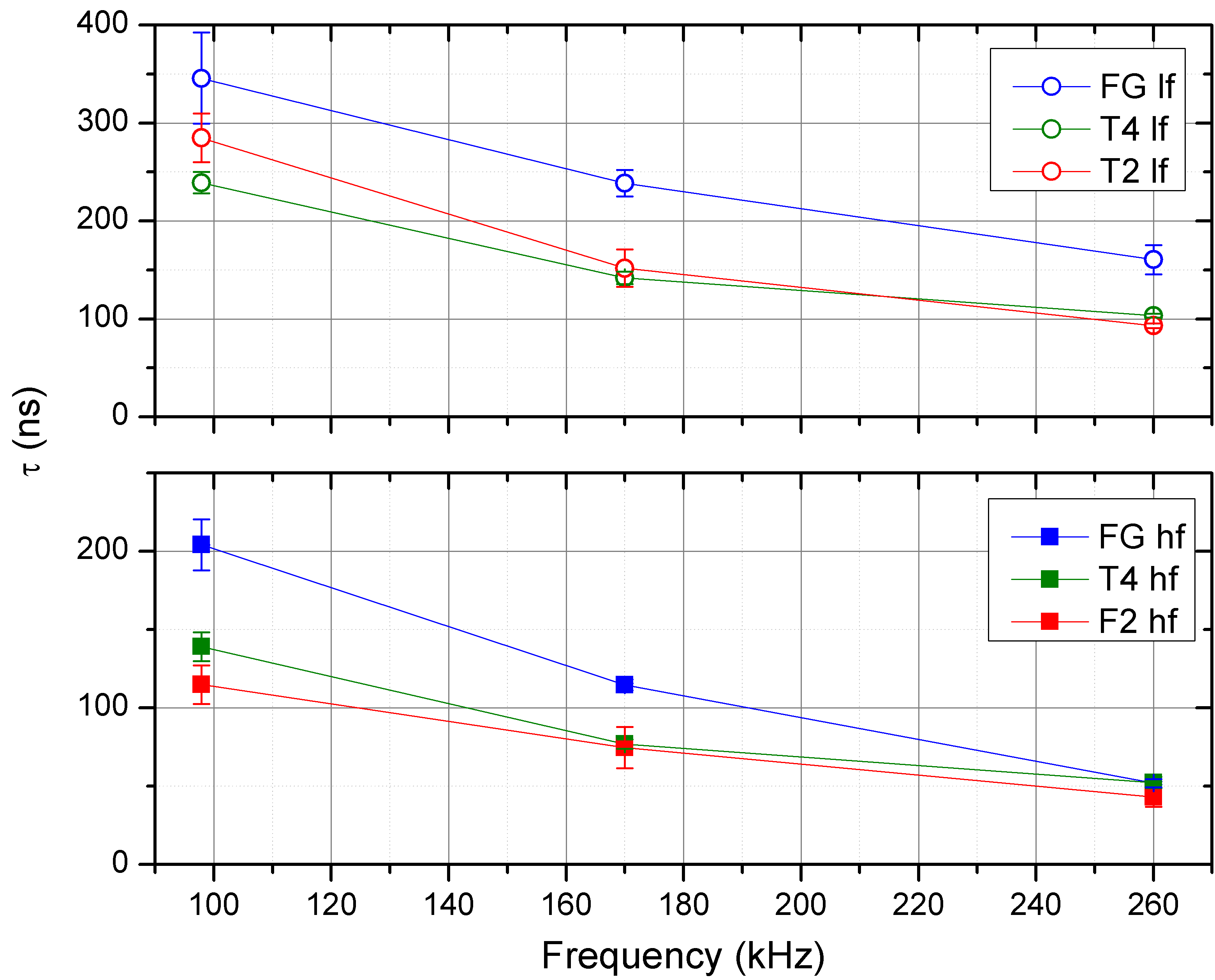}
    \caption{Relaxation time $\tau$ vs. field frequency for FG (blue), T2 (red) and T4 (green) for high (hf=54 kA/m, bottom panel) and low (lf=17.4 kA/m, top panel) field amplitudes.}
    \label{tau}
\end{figure}

Figure \ref{tau} shows the mean relaxation time $\tau$ for each sample at the three measured frequencies for low (17.4 kA/m) and high (54 kA/m) field amplitude.\\ 
Results show a reduction in $\tau$ values with frequency for all samples. While tumor samples show compatible results, FG values are slightly higher though still showing the same tendency. 

\section{Discussion}
Radiofrequency magnetic cycles of an agarose ferrogel (FG) containing Fe$_3$O$_4$ nanoflowers (NF) and of three portions of a mice tumor injected with the same particles were obtained at six frequency-amplitude field conditions. The main goal of these experiments was to compare the behavior of the NF in a typical solid medium model as the FG, with the behavior in the actual application medium represented by the tumor.\\

The portion of the tumor closest to the skin (T1) showed  the lower Fe content, compatible with the migration of the smaller MNPs away from the injection site. Iron concentration of the next tumor portion in order from the surface (T2) was similar to the concentration in the FG, while the concentration in the deeper portion of the tumor (T4) was several times higher.\\ 
For low field amplitude (17.4 kA/m) NF cycles were compatible between tumor samples showing a linear magnetic response with mostly elliptical shape. On the other hand, FG cycles depart from the linear response theory showing a more sigmoidal shape with a significant increase in uneven harmonics, similar to the results reported by Vandendriessche \textit{et al} for superparamagnetic Langevin magnetization in MNP \cite{vandendriessche2013magneto}. Coercive field is systematically higher for FG at all frequencies. SAR values are identical for both tumor portions and higher for the FG. For high field amplitude (54 kA/m) the shape of the cycles is similarly non linear for all samples even though the coercive field remains systematically higher for the FG. SAR values are again identical for both tumor portions and consistently higher for the FG at all frequencies. These SAR results are compatible with the reported by Coral \textit{et al} for \textit{ex vivo} tumors with the same batch of magnetic nanoflowers.\cite{coral2018nanoclusters}\\
Additionally, the mean relaxation time $\tau$ of the nanoparticles was determined from the cycles. The results again show compatible values between tumor samples and slightly superior values for the FG at all field conditions except for (260 kHz; 54 kA/m) where the values are compatible between all samples. $\tau$ values also show a reduction with frequency that require further investigation.\\

All these results suggest that the mobility of the NF in the FG is not the same as in the tumor tissue due to differences in the interaction between the particles and the medium. The striking similarity in the results for T2 ([Fe]=7.7(2) g/L) and T4 ([Fe]=30.9(5) g/L) indicates that the difference with FG ([Fe]=4,97(2) g/L]) is not related to the concentration of MNPs.\\ 
Thus, the validity of the characterization of nanoparticle magnetic RF response in gel matrix as a model for application environment is limited and shouldn’t be surprising to obtain different power dissipation for magnetic nanoparticles in these two media.\\ 

The results shown in this work are just a sample of the characterization features that ESAR technique allows. The possibility of obtaining the actual RF magnetic cycles of the particles inside any kind of medium opens the door to characterization far beyond the simple evaluation of dissipated power. The evaluation of the harmonic components of the time depending magnetization and the determination of the mean relaxation time for example, will be very useful in order to obtain a deeper understanding of the behavior of the particles in the application environment.

\section*{Acknowledgments}
     The authors thank CONICET and UNLP of Argentina for financial support through grants PIP 0720 and PICT-2018-3240. The group would also like to thank Mr. Pablo Mereles and Mr. Oscar \textit{Lito} Pilche for his
continuous technical assistance and advice for our experiments.

\end{document}